\documentclass[11pt,a4paper]{article}
\usepackage{jcappub}
\usepackage{graphicx}
\usepackage{epsfig}
\usepackage{bm}% bold math
\usepackage{amsfonts}
\usepackage{subfigure}
\usepackage{times}
\usepackage{natbib}
\usepackage[normalem]{ulem}

\def\lapp{\mathrel{\rlap{\raise.5ex\hbox{$<$}}
                    {\lower.5ex\hbox{$\sim$}}}}
\def\gapp{\mathrel{\rlap{\raise.5ex\hbox{$>$}}
                    {\lower.5ex\hbox{$\sim$}}}}

\title{Late-time acceleration in Higher Dimensional Cosmology}
\author{Isha Pahwa,}
\emailAdd{ipahwa@physics.du.ac.in}
\author{Debajyoti Choudhury,}
\emailAdd{debajyoti.choudhury@gmail.com}
\author{T.R. Seshadri}
\emailAdd{trs@physics.du.ac.in}
\affiliation{Department of Physics and Astrophysics, University of Delhi,\\
Delhi 110007, India.}

\date{\today}

\abstract{We investigate late time acceleration of the universe in higher
dimensional cosmology. The content in the universe is assumed to exert
pressure which is different in the normal and extra
dimensions. Cosmologically viable solutions are found to exist for
simple forms of the equation of state.  The parameters of the model
are fixed by comparing the predictions with supernovae data. 
While observations stipulate that the matter exerts almost 
vanishing pressure in the normal dimensions, we
assume that, in the
extra dimensions, the equation of state is of the form $P \propto \,
\rho^{1 - \gamma}$.
For appropriate choice of parameters, a late time
acceleration in the universe occurs with $q_0$ and $z_{tr}$ being
approximately $-0.46$ and $0.76$ respectively.}


\arxivnumber{1104.1925}
\begin{document}
\maketitle
\flushbottom
\section{Introduction}
By now, several observations have confirmed that our universe is in a
phase of accelerated expansion. Further, it is also clear that this
acceleration had set in at a relatively recent time in the cosmological
calendar. Starting with observations of Type 1A 
supernovae\cite{sn1a1,sn1a2}, this feature of the
Universe is now strongly suggested by observations of cosmic microwave
background (CMB)\cite{cmb}, large-scale structures\cite{lss}, baryon
acoustic oscillations\cite{bao} and gravitational lensing\cite{weak}. 
Although observational confirmations are piling up, a convincing
theoretical framework is still lacking. Several interesting mechanisms 
have been suggested to explain this feature of the universe, such as 
cosmological constant\cite{RevModPhys.61.1}, quintessence\cite{lcdm1}, 
modified gravity\cite{RevModPhys.82.451,PhysRevD.80.043523,
springerlink:10.1134/S0021364007150027}, chaplygin 
gas\cite{Kamenshchik2001265,PhysRevD.66.043507} and many 
others. However,
these models have their own shortcomings. For example, models with a 
non-zero cosmological constant need a high degree of fine tuning 
\cite{lcdm1,lcdm2,lcdm3,lcdm4,lcdm5,lcdm6,lcdm7}
whereas potentials required for quintessence models are unnatural in the 
context of particle physics\cite{PhysRevLett.81.3067}. Appealing to higher 
dimensional cosmological models is another promising mechanism to explain 
this mysterious phenomenon. 
This is the line of approach we adopt in this paper.

 There are various possible constructs in the extra dimensional 
context including (but not limited to) brane world models
\cite{ED1,ED2,ED3}. 
Here, we consider a particular simple model akin to that used in Ref\cite{deshdeep}.
Whereas Ref\cite{deshdeep} invoked extra dimensions 
to solve the horizon problem in early universe,
using an anisotropic fluid residing in $ 1+D_1+D_2$ dimensions, 
we adapt the formalism to produce a late time acceleration instead.

Motivated by observations, we assume that the universe  is filled with 
a uniform density matter. However, the pressure exerted by the matter 
in the normal dimensions is different from that in the compact dimensions, 
while being isotropic within each subspace. As we will argue, 
observations severely constrain the functional dependence of the pressure on
the density. Within this constraint, however, a very simple form of the 
equation of state gives an excellent agreement with data.
Although pressureless matter would, normally, decelerate the 
expansion of the universe rather than accelerating it, it is the interplay 
with the hidden dimensions that provides the impetus for this expansion.

It should be clarified at this stage that our construct is not a
brane-world scenario and that we do not attempt to address issues such
as the hierarchy problem in the Standard Model of Particle
Physics. We, rather, make the simplifying assumption that these extra
dimensions are compactified to a scale small enough to play essentially
no direct role at the TeV scale. In this sense, the spirit is closer
to more canonical scenarios defined in dimensions larger than four (an
example could be a generic model derived from String Theory). Possible
phenomenological manifestations of such models are postponed to future
discussions.

The rest of the paper is constructed as follows. In section 2, the 
formalism of the model is developed, and the equation of state argued 
for. In the subsequent section, we present the solutions to the 
ensuing evolution equations. In Section 4, we compare the predictions 
of the model with data and infer the preferred values of the parameters. 
And, finally, we conclude in Section 5.

\section{Evolution Equations}
We start with a spacetime which has, in addition to one temporal and
three normal spatial dimensions, $D$ extra spatial dimensions. This
$1+3+D$ dimensional spacetime is described by the line element
\begin{equation}
 ds^2 = -dt^2+a^2(t)\left(\frac{dr^2}{1-k_1 r^2} + r^2
 d\Omega\right)+b^2(t)\left(\frac{d R^2}{1-k_2 R^2}+R^2 d\Omega_{D-1}
\right) .
\label{lineelement}
\end{equation}
We reserve the super(sub)script `$0$' for the time dimension.
Whereas lower-case Roman indices ($i,j = 1,2,3$) denote the normal 
spatial dimensions, upper case Roman indices denote
the extra dimensions and take the values $I,J=4,5,....., D+3$. 
Here, $D$ is a parameter which takes integral values and is to be 
fixed by comparing with observations.

In eq. (\ref{lineelement}), $a(t)$ denotes the scale factor in the normal
($3$)- dimensions and $b(t)$ represents the scale factor in the extra
dimensions. Since we consider the entire ($1+3+D$)-dimensional
universe to be homogeneous, 
the two scale factors $a$ and $b$ are
functions only of the time $t$. As is well known, the visible
universe is well described by a vanishing spatial flatness ($k_1 = 0$)
and we shall assume the situation to be so. 
For reasons of simplicity as well as symmetry with the observed sector, 
we shall assume $k_2 = 0$ as well.

For this line-element, the components of the Einstein tensor $G_{\mu
  \nu}$ (with $\mu , \nu =0,1,\dots,D+3$) read
\begin{eqnarray}
G^{0}_{0}&=& -3 D\frac{\dot{a}}{a}\frac{\dot{b}}{b}-3\frac{\dot{a}^2}{a^2}-\frac{D(D-1)}{2} \frac{\dot{b}^2}{b^2} \label{G00} \\
G^{i}_{i}&=&-2\frac{\ddot{a}}{a}-D\frac{\ddot{b}}{b}-2 D\frac{\dot{a}}{a}\frac{\dot{b}}{b}-\frac{\dot{a}^2}{a^2}-\frac{D(D-1)}{2}\frac{\dot{b}^2}{b^2} \hspace{1cm} \forall \ i \label{Gii} \\
G^{I}_{I}&=&(1-D)\frac{\ddot{b}}{b}-3\frac{\ddot{a}}{a}+3(1-D) \frac{\dot{a}}{a}\frac{\dot{b}}{b}+ (D-1)(1-\frac{D}{2})\frac{\dot{b}^2}{b^2}-3 \frac{\dot{a}^2}{a^2}\hspace{1cm} \ \forall I \label{GII} \ .
\end{eqnarray}\\
The energy-momentum tensor is assumed to be of the form
\begin{equation}
 T^{\mu}_{\nu} = diag{(-\rho,P_a,P_a,P_a,P_b,\dots,P_b)} \label{Tmunu}
\end{equation}
where $\rho$ is the energy density of the fluid and $P_a$ ($P_b$) is
the pressure exerted in normal (extra) dimensions. This form of
energy momentum implies that there is isotropy within the subspace associated
with the normal dimensions and also within the orthogonal subspace spanned
by the extra dimensions. However, the pressures in the two subspaces are 
different. Note that it is the observed large-scale isotropy of the 
universe that prompts one to consider an isotropic matter distribution. 
No such restriction applies to the pressure exerted in the extra 
dimensions, and thus, we could as easily have considered more elaborate 
structure for $T^I_J$. However, other than adding more freedom to the 
model, this would not have resulted in any particular qualitative 
improvement to the scenario. Hence, we desist from adopting such a 
course and adopt eq. (\ref{Tmunu}).

The fact that $T^{\mu}_{\nu}$ needs to be
divergenceless ($T^{\mu}_{\nu ; \mu}=0$) implies
\begin{equation}
\frac{d}{dt}(\rho a^3 b^{D})+P_a b^{D}\frac{d}{dt}a^3
+P_b a^3 \frac{d}{dt}b^{D}=0 \ .
\label{continuity}
\end{equation}
In standard (3-dimensional) cosmology, the constituents of the
universe today are dark energy (essentially in the form of a
cosmological constant), dark matter and baryons. (The radiation energy
density has substantially redshifted and hence, has negligible
contribution to the energy density of the universe.) Of these, both
dark matter and baryons are very well approximated by pressureless
matter.
Indeed, the nature of the dark matter can be inferred 
very well as compared to the dark energy.
Consequently, we
start by assuming that the matter does not exert any pressure in the
three visible directions and $P_a=0$. On the other hand, in the extra
dimensions, it does exert a pressure of the form $P_b =P_b(\rho)$. 
Note that this implies that it is a {\em single fluid}
that exerts such an anisotropic pressure.  A mechanical analogy would
be that of gas molecules filling a space but constrained to move only
along a subspace. We could, of course, have adopted a scenario with
two fluids, each inhabiting a subspace. However, this would only have
increased the degrees of freedom in the theory without adding
qualitatively to our understanding. Hence we desist from doing this,
although such a course of action may well be necessitated when one
attempts to construct a microscopic theory.

The dark energy is, thus, directly `visible' 
only in the extra dimensions. The 
effect (late time acceleration) in the normal dimensions is through 
the evolution of extra dimensions. An infinite variety of equation of 
states for this dark energy are possible. For simplicity, we will assume 
a monomial form, viz. $P_b=w_b \rho$ with
\begin{equation}
w_b=\frac{w}{\bar{\rho}^\gamma} \ ,
\label{eqofstate}
\end{equation}  
where $w$ and $\gamma $ are parameters of the model to be chosen so as
to reproduce the observational data and $\bar{\rho}$ is defined in eq. \ref{rhoc}. 
Once again, this choice 
(reminiscent of a generalized Chaplygin gas~\cite{Jackiw:2004nm}) 
also serves
to minimize the number of free parameters in the theory.

The aforementioned energy momentum tensor governs the evolution of the 
1+3+D dimensional spacetime.
Let us now consider the Einstein equations, namely
\begin{equation}
G^{\mu}_{\nu}=\kappa T^{\mu}_{\nu} \label{GmunuTmunu},
\end{equation}
with $\kappa=8\pi G$.
The `00' component can be expressed as
\begin{equation}
 \frac{\dot{a}^2}{a^2}+D\frac{\dot{a}}{a}\frac{\dot{b}}{b}+\frac{D(D-1)}{6}
\frac{\dot{b}^2}{b^2} = \frac{8\pi G \rho}{3} \ .  \label{adotbya}
\end{equation}
Solving for ${\dot a}/a$ and bringing it to a form close to the familiar form
of the FRW equations, we have
\begin{equation}
 \frac{\dot{a}^2}{a^2} = \frac{8 \, \pi \, G \, \rho}{3} + 
\frac{D \, (2D+1)}{6} \, \frac{\dot{b}^2}{b^2} \mp  \frac{D}{2} \, 
\frac{\dot{b}}{b} \, \sqrt{\frac{D \, (D+2)}{3} \, \frac{\dot{b}^2}{b^2} + 
\frac{32 \, \pi \, G \rho}{3}} 
\end{equation}
%\hlite{  
In the absence of the last two terms on the R.H.S., this
  equation would, understandably, reduce to the standard
  form. In other words, the $\dot b / b$ dependent terms act
  as an effective dark 
  energy source\footnote{It is worth pointing out that, in models of non-minimal
coupling, terms that are linear in ${\dot a}/a$ do appear on the right
hand side of
the FRW equations as in eq. (\ref{adotbya})}. 

Since the scale factor, $b(t)$, corresponding to the extra dimensions
enters the equations only through ${\dot b}/b~(=d{\ln}b /dt)$, what is
relevant for the evolution of $a(t)$ is not the absolute value of $b$
but only the ratio by which $b(t)$ changes with time. This is because
in our model, spatial curvature in the hidden dimension $k_2=0$.  In
other words, it is not the size of the hidden world that matters, but
its fractional rate of change (compression or expansion). It is this
rate of compression of the extra dimensions that effectively acts 
like a dynamical dark energy source for the visible universe.
Note that, for a non-zero $\dot b / b$, the scale-factor in our world,
$a(t)$, evolves non-trivially even in the absence of any matter ($\rho
= 0$). This is not unexpected, because gravity does couple the two
subspaces and the contraction (expansion) of one can lead to the
expansion (contraction) of the other.  To be specific, $\rho = 0$
leads to a power-law evolution of the two scale-factors (the exponents
being determined by $D$) wherein one of them increases with time and
the other decreases.

  As is well known, usual ($1+3$)-dimensional deSitter
  cosmologies admit both expanding and contracting solutions. We choose
one of the solutions, namely, the expanding one, because we observe that
the universe is expanding and not contracting. The
  situation here is a little more subtle.  As far the `00' component
  of the Einstein equations goes, there is still a generalized
  symmetry of the form $(\dot a, \dot b) \leftrightarrow (- \dot a, -
  \dot b)$. However, as the form of the other components of $G_{\mu
    \nu}$ shows, this symmetry is not manifest. A consequence of this
  and the preceding discussion
  is that a pressureless fluid (even in a higher-dimensional
    world) would not admit `late time
  acceleration', although uniformly accelerating/decelerating
  solutions are possible\cite{Townsend}.

\section{Cosmological Solutions}
The evolution of the universe is governed by the Einstein equations 
along with eq. (\ref{continuity} \& \ref{eqofstate}). Not all of these 
are independent, though. For example, using the constraint equation 
($G_{0}^{0}=8\pi G \rho$) and the continuity equation ($T^{\mu}_{\nu ; \mu}=0$)
we may eliminate $\ddot b$. Before we do so, it is convenient to  
rescale the variables in terms of dimensionless quantities, namely
\begin{equation}
\begin{array}{rcl c rcl}
t &\equiv& \displaystyle \frac{\tau}{H_0}
& \qquad \quad & 
A' &\equiv& \displaystyle 
    \frac{a'}{a}=\frac{\dot{a}}{a \, H_0}
 \\[2ex]
\bar{\rho} &\equiv& \displaystyle \frac{\rho}{\rho_c}  \label{rhoc}
& & 
B' &\equiv& \displaystyle \frac{b'}{b}=\frac{\dot{b}}{b \, H_0} \ ,
\end{array}
\end{equation}
where $1/\rho_c = 8 \pi G / (3 H_0^2)$ and primes denote derivative with 
respect to $\tau$. In terms of these variables, the equations of motion 
now read
\begin{equation}
\begin{array}{rcl}
0& = & \displaystyle
   (D+2) \, A''+3(D+1) \, A'^2 + \frac{D \, (1 - D)}{2} \, B'^2 
       + D \, (D-1) \, A' \, B' + 3 \, D \, \Omega_0 \, \bar{\rho} \, w_b
\\[2ex]
{\bar\rho}' &=&  \, -\bar{\rho}\, \left[ 3 \, A'+ D \, (1+w_b) \, B' \right] 
\\[2.5ex]
B'&=& \displaystyle (D-1)^{-1} \, \left[ 
    - 3 \, A' \pm \sqrt{3 \, D^{-1} \, \left\{ (D+2) \, A'^2
          +6 \, \frac{(D-1)}{D} \, \bar{\rho}\Omega_0 \right\} }
\, \right] \ .
 \end{array}
\label{Grho}
\end{equation}
\begin{figure}[!htbp]
\centering
% \subfigure[]
{
\includegraphics[width=3.15in,height=2.1in]{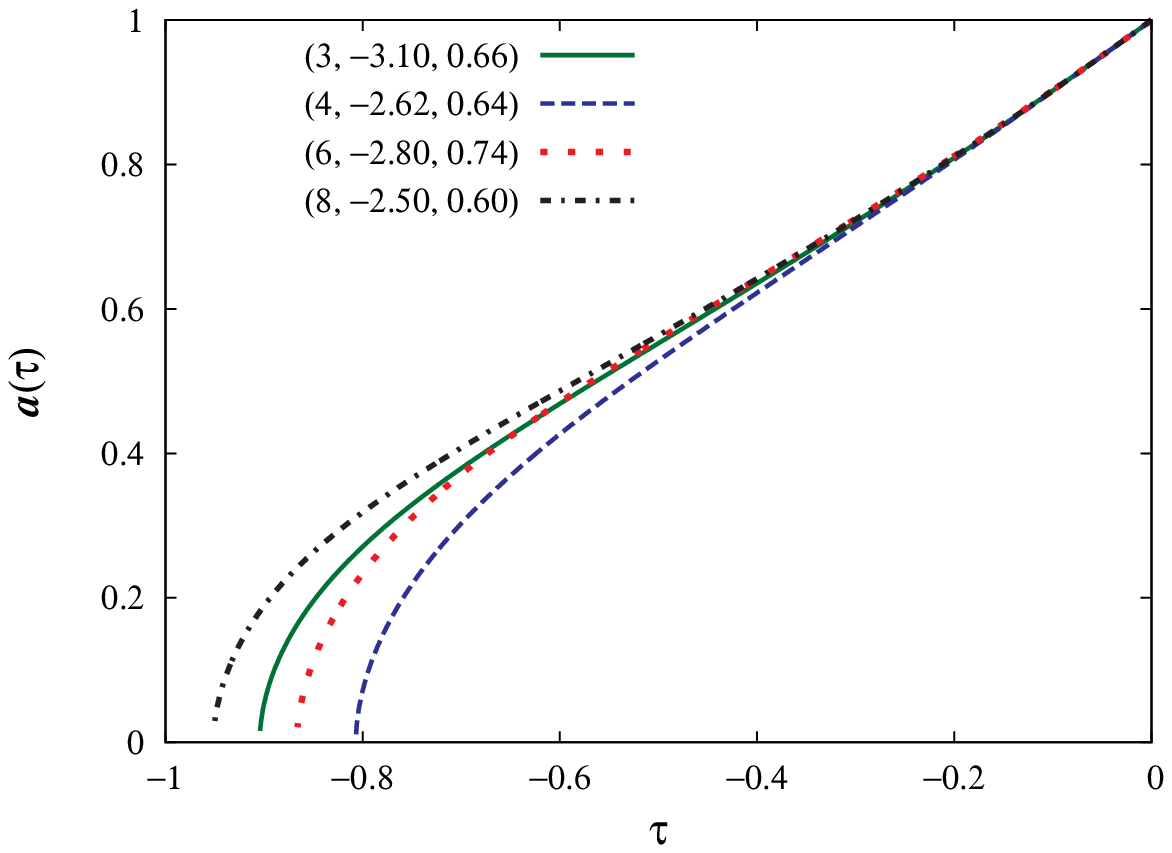}
}
% \subfigure[]
{
\includegraphics[width=3.15in,height=2.1in]{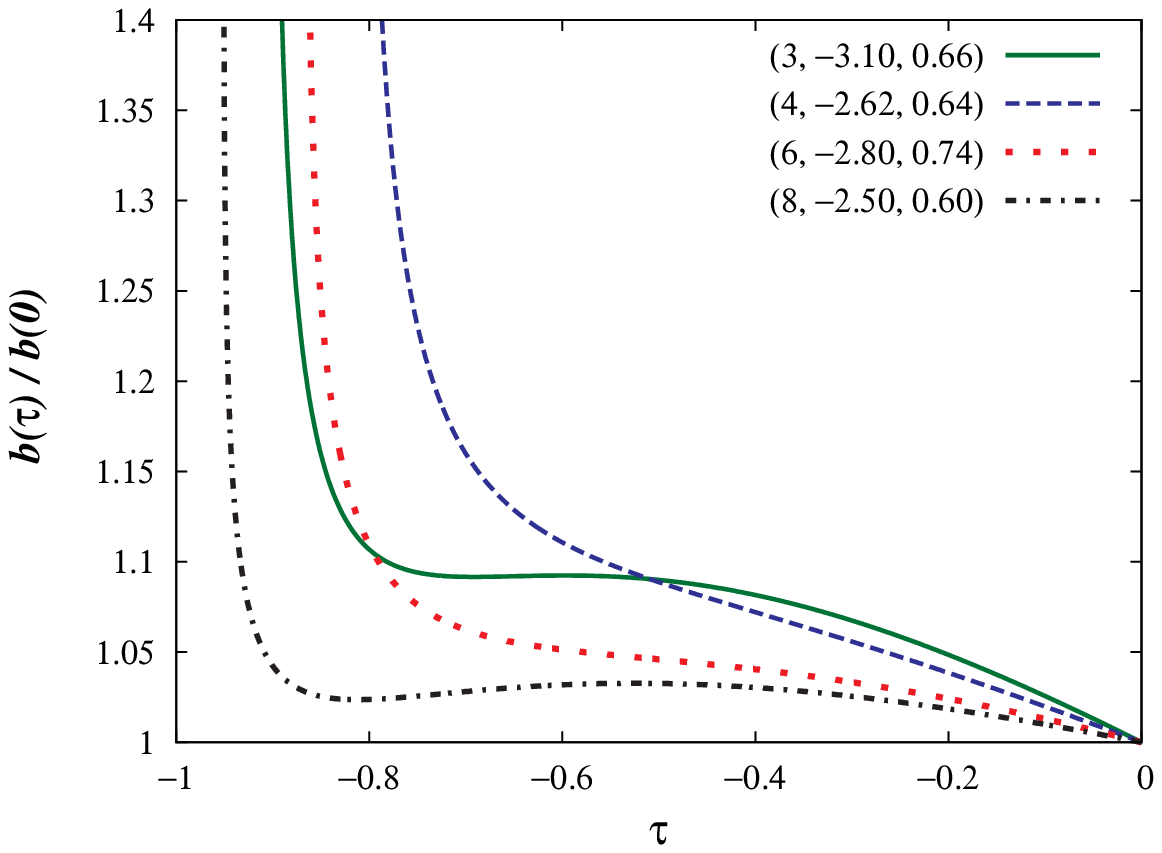}
}
\caption{Behaviour of the scale factors $ a(\tau) $ 
  [left panel] and $ b(\tau)$ [right panel] with the rescaled time $\tau$. 
  The numbers in the parentheses refer to $(D, w, \gamma)$. }
\label{fig:ab}
\end{figure}
These are two first order differential equations in 
$A'$ and $\bar \rho$ with the last of the three 
being an algebraic relation. Note that the Einstein equations 
can only determine $A'(\tau)$ and $B'(\tau)$ and 
not the scale factors themselves, a situation 
exactly analogous to the ($1+3$)-dimensional case. 
To solve eq. (\ref{Grho}) for $A'(\tau)$ and $\rho(\tau)$, 
we require two initial conditions. Since
we know the conditions in the present universe relatively
precisely, we will prescribe the conditions today.
In other words, with $\tau=0 $ referring to the 
present epoch, we evolve these equations back in time 
with the following `initial' conditions
\begin{equation}
\begin{array}{rclcrcl}
\displaystyle 
\left. \frac{\dot{a}}{a}\right|_{\tau=0}& = & H_0 & \Longrightarrow &
  A'|_{\tau=0}& = &1  
\\[3ex]
\displaystyle 
\left.\frac{\rho}{\rho_c} \right|_{\tau=0}& = & 1  
& \Longrightarrow &  \bar{\rho}|_{\tau=0} & = & 1 \ .
\end{array}
\end{equation}
We may now numerically solve the two coupled first 
order differential equations. Two such solutions exist, one for 
each sign in the last of eq. (\ref{Grho}). We reject here the branch
with the `$-$' sign as it leads to an accelerated expansion for all times 
rather than a transition from a decelerated phase to an accelerated one.

The model is characterized by three parameters namely $  D,\gamma$ and $w$. 
In figure \ref{fig:ab}, we present the solutions for the two scale factors 
for some representative values of these parameters. For ease of comparison, 
we have rescaled the solutions
\footnote{This is not to say that the two scale factors are indeed the 
same in the present epoch, but reflects the fact that the scale factors 
are arbitrary upto a constant.}
so that
\[
\begin{array}{rclcrcl}
a|_{\tau=0}&= &1 & \qquad \Longrightarrow \qquad & A|_{\tau=0}& = & 0  \\
b|_{\tau=0}& =&1 & \qquad \Longrightarrow \qquad & B|_{\tau=0}& = & 0 \ . 
\end{array}
\]

We have a whole class of solutions in which $a$, the scale factor for
our universe starts from 1 at $ \tau=0 $ and decreases monotonically
for negative values of $ \tau $. A word of caution is in place here.
Since we have neglected radiation completely, the equations are valid
only as long as the universe is matter dominated. The redshift 
corresponding to matter-radiation equality is about
$z_{eq} \sim 2.9 \times 10^4 \Omega_{m}h^2$. For the measured values
of $\Omega_m (\sim 0.27)$ and $h (\sim 0.72 )$, we have $z_{eq} \sim 3000$.
In fact, even before we go as far back as $z = z_{eq}$, the 
approximation breaks down as the radiation density can no longer be 
neglected. We have checked, though, that the inclusion of the 
radiation component does not change the evolution drastically 
for $z \gapp z_{eq}$. And, since we are primarily interested in 
the evolution of the universe in relatively recent times, the 
inclusion of radiation does not affect the results in any discernible way.

As can be easily discerned, the relative evolution in 
$b(\tau)$ is small. In the time interval that $a(\tau)$ has increased 
by nearly a factor of 3000, $b(\tau)$ has decreased by $\sim 29\%$. 
While the opposing signs of the evolution was predicted 
even in a matter-free universe (see discussion in the preceding 
section), the large difference in the magnitude of the evolution is 
but a consequence of the difference in the pressure exerted by matter
in the two worlds. 

The epoch of matter--radiation equality, $\tau_{eq} \equiv \tau(z = z_{eq})$, 
has a considerable dependence on the parameter choice. 
For a given value of $D$ and $\gamma$, a smaller $w$ shifts 
$\tau_{eq}$ further into the past [for example, 
\begin{figure}[!h]
\centering
\includegraphics[width=2.9in,height=2.1in]{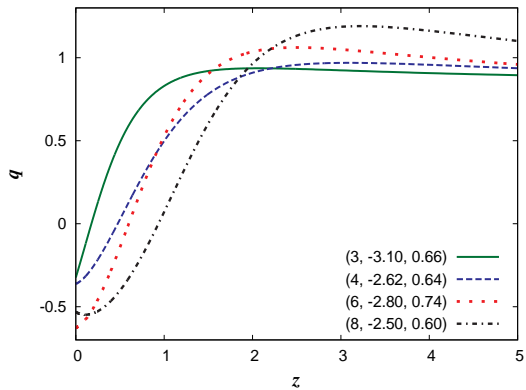}
\caption{Evolution of the deceleration parameter $q$ with the redshift $z$.
          The numbers in the parentheses refer to $(D, w, \gamma)$. }
\label{fig:decel}
\end{figure}
with $(D, \gamma)=(6, 0.59)$,
we have $\tau_{eq} \approx -0.95,-0.97,-0.99$ for $w=-2.58,-2.6,-2.62$]
thus increasing the present-day age of the universe. 
Similarly, for  a given value of $D$, and $w$, as we decrease $\gamma$, 
once again $\tau_{eq}$ shifts further into the past
[$(D, w) = (6,-2.8$) leads to $\tau_{eq} = -0.91,-0.88,-0.86$ 
for $\gamma=0.7,0.72,0.74$]. 
Note, though, that the ranges of parameters are restricted (and correlated). 
An arbitrary set would tend to destroy either late time acceleration or 
shift $\tau_{eq}$ to unacceptable values.

 As figure \ref{fig:ab} shows, the curves for $a(\tau)$ have a slight 
upward concavity for $\tau \gapp -0.3$. This is but a reflection of late 
time acceleration. Prior to this epoch, the 
universe was in a decelerated phase, as attested to by the prominent upward 
convexity at $\tau \lapp -0.5$ . This becomes clearer when we plot 
the deceleration parameter $q$ as a function of the redshift (see 
figure \ref{fig:decel}). 

\section{Observational Constraints}
 Having established that the model, for some choice of parameters, 
does lead to correct late time acceleration, we now seek to confront it with 
other observational data. The most important such data relates to Type Ia 
supernovae. The very comprehensive Union2 data set~\cite{union} lists the 
distance modulus $\mu$ as well as the redshift for 557 such supernovae. As 
the distance modulus $\mu \equiv  5 \, \log d_L + 25$ is nothing but a
rephrasing of the luminosity $d_L(z)$, defined through
\begin{equation}
 d_L(z) = (1+z) \int\limits_{0}^{z} \frac{dz'}{H(z')} \ ,
\label{dl}
\end{equation}
we, then, need to calculate $d_L(z)$, given our determination of 
$H(z)$ for a particular choice of parameters. We define a $\chi^2$--test\
through
\begin{equation}
\chi^2 (D,\gamma,w) = \sum_{i=0}^{i=n} 
   \frac{\left[ \mu_{obs}(z_i)-\mu_{th}(D,\gamma,w;z_i)\right]^2}{\sigma_i^2} 
\ ,
\end{equation}
where $\mu_{th}$ defines the value expected in our model for a 
particular choice of parameters, whereas $\mu_{obs}$ and $\sigma$ are the 
observational value and the associated root-mean-squared error. We may,
now, determine the best-fit value of the parameters by minimizing the 
$\chi^2$. 
In table \ref{table1}, we list such best fit values for 
some choices of $D$. The results are analogous for other choices. 
Note that, with an increase in $D$, both $|w|$ and $\gamma$ decrease. 
This is quite understandable as the ensuing smaller extra-dimensional 
pressures would now have an enhanced effect in the normal 
world owing to the larger effective coupling between $\dot a(t)$ and 
$\dot b(t)$. Thus, if we were to admit very large $D$ values, without 
any concern for the microscopic theory, the fluid would tend to a 
normal one. 
\begin{center}
\begin{table}[!htpb]
\centering
\begin{tabular}{|c|c|c|c|c|c|c|}
\hline
$D$ & $\chi^2_{min}$ & $w$ & $\gamma$ & $q_0$ & $z_{tr}$\\
\hline\hline
$2$ & $538.92$ & $-3.71^{+0.25}_{-0.63}$ & $0.72^{+0.23}_{-0.08}$ & $-0.49^{+0.10}_{-0.26}$ & $0.75^{+0.23}_{-0.27}$\\
\hline
$3$ & $539.18$ & $-3.10^{+0.18}_{-0.55}$ & $0.66^{+0.25}_{-0.07}$ & $-0.47^{+0.09}_{-0.27}$ & $0.78^{+0.22}_{-0.32}$\\
\hline 
$6$ & $539.33$ & $-2.50^{+0.13}_{-0.47}$ & $0.59^{+0.27}_{-0.06}$ & $-0.44^{+0.08}_{-0.29}$ & $0.77^{+0.20}_{-0.31}$\\
\hline 
$10$ & $539.42$ & $-2.31^{+0.12}_{-0.38}$ & $0.57^{+0.14}_{-0.06}$ & $-0.46^{+0.08}_{-0.36}$ & $0.79^{+0.19}_{-0.33}$\\
\hline 
$35$ & $539.44$ & $-2.06^{+0.35}_{-0.11}$ & $0.53^{+0.26}_{-0.05}$ & $-0.45^{+0.09}_{-0.26}$ & $0.79^{+0.20}_{-0.33}$\\
\hline 
$100$ & $539.74$ & $-1.98^{+0.07}_{-0.38}$ & $0.51^{+0.28}_{-0.05}$ & $-0.43^{+0.06}_{-0.30}$ & $0.80^{+0.19}_{-0.34}$\\
\hline
\end{tabular}
\\[1ex]
\caption{The values of $w$ and $\gamma$ corresponding to the 
  best fit for a given choice of $D$. The error bars correspond to 
  the projections 
  of the 95\% C.L. ellipses on the two axes. Also shown are the 
  corresponding values of $q_0$ and $z_{tr}$. }
\label{table1}
\end{table}
\end{center}

As the $\chi^2$--values listed in table \ref{table1} show, the fits are excellent. 
To further compare the shape of the theoretical spectrum with the Union2 data
set, we also performed a Kolmogorv-Smirnov test. For each of the 
cases the K-S statistic was found to be smaller than $1.8 \times 10^{-3}$ reflecting 
an extremely good fit. In fact, so good are the fits, 
that the current 
data is unable to differentiate between these
choices of parameters. 

\begin{figure}[!htbp]
\vspace*{-8ex}
\centering
% \subfigure[D=2]
{
\includegraphics[width=4.0in,height=4.0in]{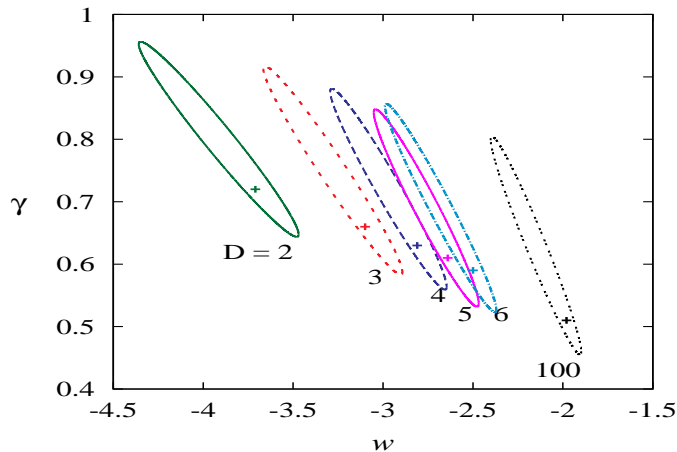}
}
\vspace*{-60pt}
\caption{$95\%$ C.L. contours in the $\gamma-w$ plane for different 
values of $D$. The points represent the best fit values for each $D$.}
\label{contour}
\end{figure}
 Since $D$ assumes only discrete values, we refrain from treating it as a
free parameter for the rest of the analysis. Rather, for a given $D$, we 
consider the $\gamma$-$w$ plane as a two-dimensional parameter space. 
We may, then, attempt to define $95\%$ C.L. contours in this plane 
by considering $\Delta \chi^2$. These are displayed
in figure \ref{contour}.
As is evident, there is a strong negative correlation between the 
two parameters. Note, furthermore, that positive $w$ (hence, positive pressure)
is essentially ruled out. Similarly, integral values of $\gamma$ are 
also essentially ruled out. 
Further, this is true even if we consider values of 
$D$ far larger than those preferred by microscopic theories of 
high energy physics.
As can be deduced from table \ref{table1}, each of the marked points 
in figure \ref{contour} denotes essentially a global minimum of 
$\chi^2$, with the position of minima getting increasingly closer as one increases 
$D$ arbitrarily.

Once $a(\tau)$ and, hence, $H(z)$ is determined in a model, one may 
also calculate both the present deceleration parameter $q_0$ as well
as $z_{tr}$, the redshift corresponding to the epoch of transition from 
the decelerated to the accelerated phase. Note that these values are 
not uniquely 
determined by the data alone as the cosmological model has a strong 
bearing on this determination. Also shown, in table \ref{table1}, are the 
values of $q_0$ and $z_{tr}$ as determined within our model.
\begin{figure}[!h]
\centering
\includegraphics[width=2.9in,height=2.1in]{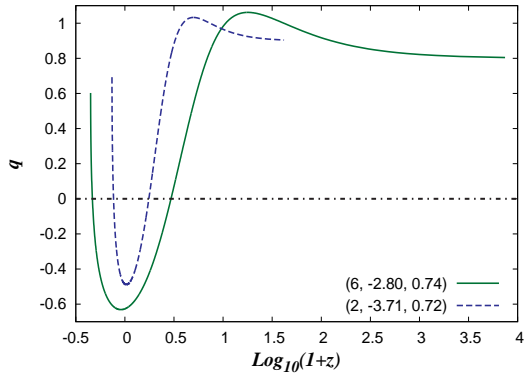}
\caption{Evolution of the deceleration parameter $q$ with the redshift.
          The numbers in the parentheses refer to $(D, w, \gamma)$. }
\label{logzq}
\end{figure}
\section{Conclusions and Discussion}
Several approaches have been explored in the literature
to arrive at a late-time acceleration of the Universe.
However, none of the models are completely satisfactory. Most models rely 
on a mysterious type of matter (Dark Energy) which gives an effective 
repulsive gravity which is supposed to provide the observed accelerated 
expansion. This mysterious source term is perceived only through its 
effect on cosmological expansion.

In this paper, we have followed an approach using higher dimensions.
The mysterious behaviour of matter ({\it a la} the Dark
Energy) is manifested directly only in the extra
dimensions. The accelerated expansion of the scale factor in normal
dimensions is produced only indirectly by the source term through its
effect on the extra dimensions.  Thus, in our model, the issue of
invoking strange properties for the dark energy fluid does not arise
as it exhibits its unusual property only in the extra dimensions.

The particularly simple scenario we consider here is able to produce
the requisite late time acceleration. The matter content acts as a
pressureless gas in the normal dimensions and has a monomial equation
of state as far as the extra dimensions are concerned. Just this
simple ansatz leads to not only a late time acceleration, but also to
a very moderate contraction of the extra dimensions since the epoch of
radiation-matter equality although the normal dimensions have expanded
by a factor of $\sim 3000$ in the same time period. With the size of
the extra dimensions hardly changing, the role of any fields confined
to the extra-dimensions as also those of any possible Kaluza-Klein
towers of the SM fields in the low energy limit has remained
essentially unaltered in this epoch.

For a significantly wide range of parameters, the model shows
excellent agreement with the observational data on Type Ia supernovae,
as attested to by both a low $\chi^2$ per degree of freedom as well as
the Kolmogorov-Smirnov statistic. We still need a negative pressure
nonetheless, albeit limited to the extra dimensions. Whether a more
complicated scenario, involving a non-isotropic extra dimensional
subspace and/or multiple fluids, obviates this restriction is yet to
be seen.
  
An interesting feature of our model is that the present phase of
accelerated expansion is, generically, a transient one. In
figure \ref{logzq}, we plot $q(z)$ against $\log_{10}(1+z)$. Note that the
future is defined by $\log_{10}(1+z)<0$.  For example, $(D, w, \gamma) = (2,
-3.71, 0.72)$, namely one of our best fit points, the universe will
transit to a decelerating phase at $z = -0.23$. Similarly, for $(D, w,
\gamma) = (6, -2.80, 0.74)$, which is somewhat away from a best fit
(but within the 95\% C.L region), this transition would occur at $z =
-0.28$.

One aspect which we have not dwelt with in this paper is the rate of
growth of perturbations. While a detailed analysis is beyond the scope
of this paper and would be addressed elsewhere, let us make a few
comments.  With the dynamics presented here being the dominant driving
mechanism post matter-radiation equality, changes in structure
formation would be confined to this epoch. In fact, the quantum of
difference from the $\Lambda$-CDM scenario would be of the same
approximate size as in a large class of theories with a dynamical Dark
Energy source.  This is supported by the fact that $q_0$
in this model (see table \ref{table1}) agrees (within error bars) with $-0.6$, the
value it assumes in the standard $\Lambda$-CDM scenario. Indeed, a
careful analysis of this aspect could prove of value in further
narrowing down of the parameter space, with larger $D$ being less
preferred.

\acknowledgments{The authors thank the anonymous referee for
  important suggestions that led to an improvement in the presentation.
  DC thanks the Department of Science and Technology, India
  for assistance under the project DST-SR/S2/HEP-043/2009.  IP
  acknowledges the CSIR, India for assistance under grant
  09/045(0908)/2009-EMR-I. Authors acknowledge facilities provided by 
Inter University Center For Astronomy
and Astrophysics, Pune, India through IUCAA Resource
Center(IRC) at Department of Physics and Astronomy,
University of Delhi, New Delhi, India.}

\end{document}